\documentclass[prl,twocolumn,twoside,preprintnumbers,superscriptaddress,nofootinbib]{revtex4}

\usepackage{amsmath}
\usepackage{graphicx}
\usepackage{dcolumn}
\usepackage{bm}
\usepackage[hyperfootnotes=false]{hyperref}
\usepackage{xspace}

\newcommand{\Order}{\mathcal{O}}
\newcommand{\beq}{\begin{equation}}
\newcommand{\eeq}{\end{equation}}
\newcommand{\diff}{\text{d}}
\newcommand{\Fpi}{F_\pi}

\newcommand{\mpii}{M_{\pi^0}}

\newcommand{\F}{\mathcal{F}}
\newcommand{\MeV}{\,\text{MeV}}

\begin{document}

\title{Remarks on higher-order hadronic corrections to the muon $\boldsymbol{g-2}$}

\author{Gilberto Colangelo}
\author{Martin Hoferichter}
\author{Andreas Nyffeler}
\affiliation{Albert Einstein Center for Fundamental Physics,
Institute for Theoretical Physics,\\ University of Bern, Sidlerstrasse 5, CH--3012 Bern, Switzerland}

\author{Massimo Passera}

\affiliation{Albert Einstein Center for Fundamental Physics,
Institute for Theoretical Physics,\\ University of Bern, Sidlerstrasse 5, CH--3012 Bern, Switzerland}
\affiliation{Istituto Nazionale Fisica Nucleare, Sezione di Padova, I--35131 Padova, Italy}

\author{Peter Stoffer}
\affiliation{Albert Einstein Center for Fundamental Physics,
Institute for Theoretical Physics,\\ University of Bern, Sidlerstrasse 5, CH--3012 Bern, Switzerland}

\begin{abstract}
Recently, it was shown that insertions of hadronic vacuum polarization at $\Order(\alpha^4)$ generate non-negligible effects in the calculation of the anomalous magnetic moment of the muon. This result raises the question if other hadronic diagrams at this order might become relevant for the next round of $g-2$ measurements as well. In this note we show that
a potentially enhanced such contribution, hadronic light-by-light scattering in combination with electron vacuum polarization, is already sufficiently suppressed.
\end{abstract}

\maketitle

\begin{figure}
\includegraphics[width=0.5\linewidth]{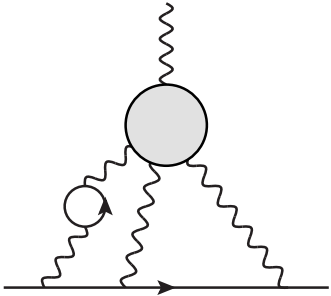}
\caption{HLbL scattering combined with lepton vacuum polarization. The grey blob refers to the HLbL amplitude and solid/wiggly lines to leptons/photons. Diagrams where the lepton loop is inserted into the other photon propagators are not shown.}
\label{fig:HLbL_NLO}
\end{figure}

In~\cite{Kurz:2014wya} the contribution of diagrams involving hadronic vacuum polarization (HVP) at $\Order(\alpha^4)$ to the anomalous magnetic moment of the muon was calculated as
\beq
a_\mu^\text{HVP, NNLO}=(12.4\pm 0.1)\cdot 10^{-11}.
\eeq
This result is significantly larger than expected if compared to the suppression of $|a_\mu^\text{HVP, NLO}/a_\mu^\text{HVP, LO}|\approx 1/70$~\cite{Hagiwara:2011af}, which would have suggested an estimate of $|a_\mu^\text{HVP, NNLO}|\approx 1.4\cdot 10^{-11}$.
This substantial enhancement of NNLO HVP diagrams immediately raises the question if other hadronic contributions at $\Order(\alpha^4)$ might be non-negligible as well. Amongst these missing $\Order(\alpha^4)$ contributions is hadronic light-by-light scattering (HLbL) combined with lepton vacuum polarization, see Fig.~\ref{fig:HLbL_NLO}. For the electron this diagram is enhanced by $\log m_\mu/m_e$ and therefore could become relevant in case of a large prefactor. In particular, taking~\cite{Jegerlehner:2009ry}
\beq
\label{HLbL}
a_\mu^\text{HLbL, LO}=(116\pm 39)\cdot 10^{-11}
\eeq
for HLbL scattering,
a suppression factor similar to $|a_\mu^\text{HVP, NNLO}/a_\mu^\text{HVP, NLO}|\approx 1/8$ would indicate a contribution of $|a_\mu^\text{HLbL, NLO}|\approx 15\cdot 10^{-11}$, of the same order as the accuracy projected for upcoming experiments~\cite{Venanzoni:2012qa,Saito:2012zz}.

The polarization due to $\ell^+\ell^-$ pairs ($\ell=e,\mu,\tau$) leads to a modification of the photon propagator by a factor
\beq
\label{vac_pol}
\Pi\big(q^2\big)=\frac{2\alpha}{\pi}\int\limits_0^1\diff x\, x(1-x)\log\bigg[1-x(1-x)\frac{q^2}{m_\ell^2}\bigg],
\eeq
with photon virtuality $q^2$ and lepton mass $m_\ell$. For illustration, we approximate the HLbL tensor by the pion-pole contribution~\cite{Knecht:2001qf}
\begin{align}
\label{amu_pole}
 a_\mu^{\pi^0\text{-pole}}&=-e^6\int\frac{\diff^4q_1}{(2\pi)^4}\int\frac{\diff^4q_2}{(2\pi)^4}
 \frac{1}{q_1^2q_2^2sZ_1 Z_2}\notag\\
 &\times\Bigg\{\frac{\F_{\pi^0\gamma^*\gamma^*}\big(q_1^2,q_2^2\big)\F_{\pi^0\gamma^*\gamma^*}\big(s,0\big)}{s-\mpii^2}T^{\pi^0}_1(q_1,q_2;p)
 \notag\\
 &\quad\!+\frac{\F_{\pi^0\gamma^*\gamma^*}\big(s,q_2^2\big)\F_{\pi^0\gamma^*\gamma^*}\big(q_1^2,0\big)}{q_1^2-\mpii^2}T^{\pi^0}_2(q_1,q_2;p)\Bigg\},\notag\\
 Z_1&=(p+q_1)^2-m_\mu^2,\qquad Z_2=(p-q_2)^2-m_\mu^2,\notag\\
 s&=(q_1+q_2)^2,
\end{align}
and kinematic factors $T^{\pi^0}_i(q_1,q_2;p)$ as given in~\cite{Knecht:2001qf} (see~\cite{Colangelo:2014dfa} for an interpretation within dispersion theory). $p$ denotes the muon momentum and $\F_{\pi^0\gamma^*\gamma^*}(q_1^2,q_2^2)$ the pion transition form factor. 
In most model calculations of HLbL the pion pole is the most important single contribution, and the full result is dominated by the sum of all light pseudoscalar mesons~\cite{Jegerlehner:2009ry}.
For the pion transition form factor we take a simple VMD model
\beq
\F_{\pi^0\gamma^*\gamma^*}(q_1^2,q_2^2)=\frac{1}{4\pi^2F_\pi}\frac{M_\rho^4}{\big(M_\rho^2-q_1^2\big)\big(M_\rho^2-q_2^2\big)},
\eeq
with pion decay constant $\Fpi=92.2\MeV$ and $M_\rho=775.26\MeV$~\cite{Beringer:1900zz}. This model for the pion transition form factor leads to
\beq
a_\mu^{\pi^0\text{-pole}}=57.2\cdot 10^{-11},
\eeq
in agreement with~\cite{Knecht:2001qf}. Modifying the photon propagators by the polarization factor~\eqref{vac_pol}, we find for the electron loop
\beq
\label{HLbL_NLO_e}
a_\mu^{\pi^0\text{-pole, NLO}}=1.5\cdot 10^{-11},
\eeq
a mere $2.6\%$ correction. In fact, from renormalization-group arguments~\cite{Lautrup:1974ic} one would have expected a suppression of
\beq
\label{suppression_factor}
3\times\frac{\alpha}{\pi}\times\frac{2}{3}\log\frac{m_\mu}{m_e}\approx 2.5\%,
\eeq
in remarkable agreement with the explicit calculation. In this estimate, the factor $3$ originates from the fact that each of the photon propagators can be renormalized, and the prefactor of the logarithm can immediately be derived from~\eqref{vac_pol} in the limit $m_\ell\to 0$.
We also checked diagrams with muon/tau loops and HVP, all of which were found to be at least another order of magnitude smaller than~\eqref{HLbL_NLO_e}.

Further $\Order(\alpha^4)$ terms involve radiative corrections to the muon line. In order to estimate the possible impact of such contributions, we compare in QED the diagrams where the HLbL amplitude is replaced by a muon loop, whose mass is sufficiently close to typical hadronic scales to serve as an indication of the order of magnitude of effects to be expected. The contribution with electron-vacuum-polarization corrections is given as a subclass to IV(a) in~\cite{Kinoshita:2004wi}, the radiative corrections to the muon line correspond to class IV(c) in~\cite{Kinoshita:2005zr}, with a ratio $-4.33/1.14\approx -4$ (similarly, class IV(d) in~\cite{Kinoshita:2005zr}, which would correspond to fully-offshell HLbL scattering, is suppressed by $0.99/4.33$). Therefore, also these remaining radiative corrections are unlikely to upset the estimate presented here. Finally, there are in principle also radiative corrections to HLbL scattering, but these are suppressed for the dominant pseudoscalar poles,\footnote{Since the exchanged meson does not carry electric charge, radiative corrections to the pion pole vanish unless the photon resolves the internal structure, which implies a suppression (for instance in a chiral counting).} so that this effect should be safely encompassed by the intrinsic uncertainty in the HLbL amplitude. Taking everything together, we obtain the estimate
\beq
a_\mu^\text{HLbL, NLO}=(3\pm 2)\cdot 10^{-11},
\eeq
where the central value follows from~\eqref{HLbL} and the suppression factor~\eqref{suppression_factor}, with uncertainties conservatively estimated from~\eqref{HLbL} and the radiative corrections as observed for the muon loop.

In conclusion, there is no evidence for any $\Order(\alpha^4)$ diagram involving HLbL significantly surpassing its naive estimate. In particular, we find that the size of potentially relevant diagrams enhanced by an insertion of an electron loop can be well estimated by renormalization-group arguments, as verified for a HLbL amplitude approximated with a $\pi^0$ pole and VMD form factors. The resulting estimate for $a_\mu^\text{HLbL, NLO}$ lies a factor $5$ below the accuracy goal of the next round of $g-2$ experiments and can therefore be presently neglected. 

{\it Acknowledgements}  
Financial support by the Swiss National Science Foundation is gratefully acknowledged.
The AEC is supported by the
  ``Innovations- und Kooperationsprojekt C-13'' of the ``Schweizerische
  Universit\"atskonferenz SUK/CRUS.''
 The work of MP was supported in part by the Italian Ministero dell'Universit\`a e della Ricerca Scientifica under the program PRIN 2010-11 and by the European Programme INVISIBLES (contract PITN-GA-2011-289442).

\newpage

\end{document}